\documentclass[twocolumn,secnumarabic,amssymb, nofootinbib, nobibnotes,superscriptaddress,showkeys, aps, prd]{revtex4-2}

\usepackage{graphicx}	
\usepackage{amsmath}	
\usepackage{amssymb}	
\usepackage{hyperref}
\usepackage{subfigure}
\usepackage{romannum}

\newcommand{\diff}{{\text{d}}}

\newcommand{\aap}{    {\it Astron. \& Astrophys.}}

\newcommand{\apjl}{   {\it Astrophys. J. Lett.}}
\newcommand{\apss}{   {\it Astrophys. Space Sci.}}

\newcommand{\mnras}{  {\it Mon. Not. Roy. Astron. Soc.}}
\newcommand{\pasp}{   {\it Pub. Astron. Soc. Pac.}}

\newcommand{\ssr}{    {\it Space Sci. Rev.}}

\newcommand{\rmxaa}{ {\it Revista Mexicana de Astronoma y Astrofsica}}

\usepackage{color}

\begin{document}

\title{Revealing Physical Properties of a Tidal Disruption Event: iPTF16fnl}

\author{Mageshwaran Tamilan}%
\affiliation{Department of Space Science and Astronomy, Chungbuk National University, Cheongju 361-763, Korea}

\author{Gargi Shaw}%
\affiliation{Department of Astronomy and Astrophysics, Tata Institute of Fundamental Research, 1 Homi Bhabha Road, Mumbai 400005, India}

\author{Sudip Bhattacharyya}%
\affiliation{Department of Astronomy and Astrophysics, Tata Institute of Fundamental Research, 1 Homi Bhabha Road, Mumbai 400005, India}

\author{Kimitake Hayasaki}%
\affiliation{Department of Space Science and Astronomy, Chungbuk National University, Cheongju 361-763, Korea}
\affiliation{Department of Physical Sciences, Aoyama Gakuin University, Sagamihara 252-5258, Japan}

\email{tmageshwaran2013@gmail.com, tmageshwaran@chungbuk.ac.kr}
\date{\today}%

\begin{abstract}
Tidal disruption event (TDE) iPTF16fnl shows a relatively low optical flare with observationally very weak X-ray emission and the spectroscopic property that the helium emission line from the source dominates over the hydrogen emission line at early times. We explore these observed signatures by calculating spectral emission lines with the publicly available code, CLOUDY. We estimate five physical parameters by fitting the observed optical UV spectra on multiple days to a theoretical model of a steady-state, slim disk with a spherical outflow. The resultant key parameters among them are black hole mass $M_{\bullet} \sim 4.43 \times 10^5 M_{\odot}$, stellar mass $M_{\star} \sim 0.46 M_{\odot}$, and wind velocity $v_{\rm w} \sim 5382.7~{\rm km~s^{-1}}$. The disk-wind model also estimates the radiative efficiency to be $0.01\lesssim\eta\lesssim0.04$ over the observational time, resulting in the disk being radiatively inefficient. In our CLOUDY model, the filling factor of the wind is also estimated to be 0.8, suggesting that the wind is moderately clumpy. We reveal that the helium-to-hydrogen number density ratio of the wind lies between 0.1 and 0.15, which is nearly the same as the solar case, suggesting the tidally disrupted star is originally a main sequence star. Because the optical depth of the helium line is lower than the hydrogen line by two orders of magnitude, the helium line is significantly optically thinner than the hydrogen line. Consequently, our results indicate that the helium line luminosity dominates the hydrogen line luminosity due to the optical depth effect despite a small helium-to-hydrogen number density ratio value.
\end{abstract}

\maketitle



\section{Introduction}

Tidal disruption events (TDEs) are transient astronomical phenomena that occurs when a star approaches a supermassive black hole (SMBH) closely enough for the SMBH tidal force to tear
apart the star \citep{1976MNRAS.176..633F,1988Natur.333..523R}. 
The tidal disruption radius is the distance of the star from the black hole below which the entire star is disrupted and is given by $r_{\rm t}= (M_{\bullet}/M_{\star})^{1/3} R_{\star}$, where $M_{\bullet}$ is the black hole mass, and $M_{\star}$ and $R_{\star}$ are the stellar mass and radius respectively \citep{1976MNRAS.176..633F}. The stellar mass fraction stripped during disruption depends on the orbital pericenter and stellar density profile \citep{2013ApJ...767...25G}. The disrupted debris returns to the pericenter with a mass fallback rate that follows $t^{-5/3}$ evolution at late times, where $t$ is the time  \citep{2009MNRAS.392..332L,2015ApJ...814..141M}. However, the mass fallback rate deviates from $t^{-5/3}$ law at early times due to the stellar density \citep{2009MNRAS.392..332L}, stellar rotation \citep{2019ApJ...872..163G} and stellar orbital eccentricity \citep{2018ApJ...855..129H,2020ApJ...900....3P,2023ApJ...959...19Z,2022ApJ...924...34C}. Subsequently, the debris stream-stream collision causes an energy dissipation, leading to the formation of an accretion disk \citep{2013MNRAS.434..909H,2016MNRAS.461.3760H,2016MNRAS.455.2253B,bonnerot_simulating_2020}. 
Considering that outflows (i.e., winds) are blowing from the formed disk or initial strong stream-stream collisional point \citep{lu_self-intersection_2020} in TDEs without relativistic jets, the photons are produced at the respective sites. Those photons are reprocessed or emitted from the disk and the photosphere of the wind, which have been observed over a wide spectral range from infrared (IR), optical, ultraviolet (UV), and X-ray wavebands.

In optical/UV TDEs, the optical/UV emission dominates at early times, and the observational X-ray flux is lower than that expected from the thermal disk spectrum. The blackbody temperatures estimated by fitting the X-ray spectra for ASAS-SN 14li,  XMMSL1 J061927.1-655311, Abell-1795, and NGC-3599 are $\sim 10^5$ K \citep{2016MNRAS.455.2918H}, $\sim 1.4 \times 10^6$ K \citep{2014A&A...572A...1S}, $\sim 1.2 \times 10^6$ K \citep{2013MNRAS.435.1904M} and $\sim 1.1 \times 10^6$ K \citep{2007A&A...462L..49E}, respectively. A part of X-ray photons emitted from the disk can escape along the vertical direction \citep{2018ApJ...859L..20D}. Those photons are sometimes observed to be weak X-ray emission in the optical/UV TDEs.The remaining X-ray photons are thought to be reprocessed to the longer optical/UV wavelengths by the outflow from the disk \citep{2009MNRAS.400.2070S,2020ApJ...894....2P}. The resultant optical/UV emission at the photosphere has a blackbody temperature ($\sim~{\rm few} \times 10^4$ K) which is an order smaller than the X-ray blackbody temperature \citep{2020SSRv..216..114R}.

In the past works of literature, the outflow from the disk is assumed to have a spherical geometry with a radial profile of the density $\rho(r) \propto r^{-2}$ \citep{2009MNRAS.400.2070S,2020ApJ...894....2P,2020ApJ...897..156U}. 
The photons are coupled with the gas within the trapping radius ($r_{\rm tr}$), where the photon diffusion time is longer than the wind dynamical time. The wind expands adiabatically for $r < r_{\rm tr}$ and while the gas is cooled through photon diffusion for $r > r_{\rm tr}$. The gas temperature reduces with $r$, and the radial profile of temperature varies in the adiabatic and diffusive regions \citep{2020ApJ...894....2P,2020ApJ...897..156U}. \citet{2023MNRAS.518.5693M} fit the optical/UV emissions from the spherical outflow emerging from a slim disk and fit the optical/UV continuum of iPTF16axa to estimate the stellar mass and wind properties such as density, temperature, and velocity by assuming a steady-state spherical outflow.
  
The spectrum of a TDE consists of the broad emission lines with blue continuum \citep{2016MNRAS.463.3813H,2017apj...844...46B,2017ApJ...842...29H}. The helium line, HeII ($\lambda$4686 \AA), and the hydrogen line, H$\alpha$ ($\lambda$6562 \AA), are the most notable emission lines observed in TDE spectra. It is often shown that the helium line luminosity dominates the hydrogen line luminosity in the spectrum \citep{2016MNRAS.455.2918H,2017ApJ...842...29H,2017apj...844...46B}. TDEs exhibit a wide range of hydrogen to helium line ratios, varying from nearly 1:1 to spectra showing only hydrogen or only helium lines. For example, TDEs PTF09djl \citep{2014ApJ...793...38A} and ASASSN-14ae \citep{2014MNRAS.445.3263H} are H-strong TDEs, while PS1-10jh \citep{2012Natur.485..217G}, PTF09ge \citep{2014ApJ...793...38A}, and ASASSN-15oi \citep{2016MNRAS.463.3813H} are He-strong TDEs with little-to-no hydrogen lines. Additionally, TDEs such as ASASSN-14li \citep{2016MNRAS.455.2918H}, iPTF16axa \citep{2017ApJ...842...29H}, and AT2017eqx \citep{2019MNRAS.488.1878N} display both helium and hydrogen lines in their spectra.

\citet{2016ApJ...827....3R} assumed an optically thick static atmosphere with fixed inner and outer radii of the atmosphere and radial density profile of $\rho(r) \propto r^{-2}$ in their radiative transfer code with a given luminosity to calculate the reprocessed spectrum. The atmosphere contains hydrogen, helium, and oxygen in solar abundances. Photons absorbed by the atoms in the atmosphere can either be line-scattered or destroyed by processes like de-excitation, photoionization, or radiative excitation. These processes continue until the photon reaches the thermalization depth, where it can escape without further absorption. This depth corresponds to the optical depth and depends on wavelength, temperature, and atomic populations. The suppression of Balmer lines, leading to the dominance of helium emission lines, is caused by optical depth effects. Also, the dominance of helium lines over hydrogen lines can be explained by the high helium abundance in the atmosphere because of the disruption of an evolved star \citep{2015mnras.454.2321s}. Recently, \citet{2023MNRAS.518.5693M} studied the spectral properties of iPTF16axa by applying the disk-wind model and the CLOUDY modeling to the optical/UV continuum and emission lines of TDE iPTF16axa. They demonstrated that the super solar abundance of He, as well as a smaller He II line optical depth, are responsible for the enhancement of helium lines over the hydrogen lines. Overall, to elucidate the physical properties of TDEs, the simultaneous application of the disk-wind and CLOUDY models, which can reproduce the observational continuum flux and emission lines, respectively, is an effective method.

iPTF16fnl was discovered on UT 2016 August 29.4 in the g and R bands by intermediate Palomar Transient Factory (iPTF) \citep{2017apj...844...46B}. iPTF16fnl is a relatively faint optical TDE located at the center of an E+A galaxy (Mrk950) at a distance of 66.6 Mpc at redshift 0.016328 \citep{{2016ATel.9433....1G},{2017apj...844...46B},{2018MNRAS.473.1130B}}. No prior activity in the host galaxy is detected with upper limits of $\sim20-21$ mag, whereas the transient is observed with a peak magnitude of $\sim$ 17 mag in the g band. The time of the g-band peak, MJD 57632.1, was taken as the reference epoch, where day 0 corresponds to this peak. iPTF16fnl is dominated by optical/UV emissions with no significant X-ray observations. The bolometric luminosity estimated using a single blackbody temperature fit on the optical/UV continuum shows a peak value of $\simeq (1.0 \pm 0.15) \times 10^{43}~{\rm erg~s^{-1}}$, which is an order smaller than the Eddington luminosity:
\begin{equation}
L_{\rm Edd} = \frac{4 \pi G M_{\bullet} c}{\kappa_{\rm es}}
=
(3.16 \pm 2.76) \times 10^{44}~{\rm erg~s^{-1}}
\end{equation}
where $G$ is the gravitational constant, $\kappa_{\rm es} = 0.34~{\rm cm^2~g^{-1}}$ is the Thomson electron scattering opacity, $c$ is the speed of light, and we adopt $M_{\bullet} = (2.138 \pm 1.87) \times 10^6 M_{\odot}$ as the black hole mass \citep{2017apj...844...46B}. The single blackbody temperature of the observed emission has an average value of $T_{\rm BB} = 19000 \pm 2000 ~{\rm K}$ and does not vary significantly. The observed bolometric luminosity estimated using a single temperature blackbody model decreases by an order of 100 in around 60 days, indicating a fast-evolving TDE. The dominance of optical/UV emissions implies that there is an atmosphere that hinders the observation of disk X-ray radiation, and the emission is from the atmosphere in the lower wavelengths. The Spectroscopic analysis of the early epoch spectra ($<$ 50 days) shows the characteristic blue continuum component along with the most prominent emission lines corresponding to broad HeII and H$\alpha$. The Helium line dominates at early times and becomes significantly weak after $\sim$ 30 days. 



In this paper, we explore the physical properties of iPTF16fnl using the same but improved methods as those used for iPTF16axa by  \citet{2023MNRAS.518.5693M}. First, we study the elemental abundance of the atmosphere and emission line luminosities by CLOUDY and compare them with the observational emission line luminosities, targeting iPTF16fnl. Next, we determine the key physical parameters of iPTF16fnl by fitting the disk-wind model to the observational continuum spectra. The paper is organized as follows. In Section~\ref{cloudy}, we estimate the helium and hydrogen line luminosities and the elemental abundances in the atmosphere by using CLOUDY, and then we compare them with the observational ones. Section~\ref{dwfit} describes the disk-wind model, in particular, the ratio of the mass outflow to fallback rates, and the radiative efficiency. The detail of the disk-wind model is summarized in Appendix~\ref{dwmodel_theory}. Section~\ref{dwres} presents the results from the disk-wind model. We discuss the results from CLOUDY and disk-wind modelings in section \ref{discuss}. Section\ref{summary} is devoted to our conclusions.

%
\section{CLOUDY models}
\label{cloudy}
%

In this section, we briefly describe our method with the numerical spectroscopic simulation code CLOUDY (c22.02) \citep{2017RMxAA..53..385F}\footnote{\url{https://www.nublado.org}}. CLOUDY models the ionization, temperature, and emission spectra of gas clouds by balancing heating and cooling processes like photoionization, recombination, and other microphysical effects that influence element ionization. The code assumes that radiation interacts with the gas, either escaping or being absorbed and re-emitted, depending on the optical depth at various wavelengths. In fact, CLOUDY simulates the thermal, ionization, and chemical structure of astrophysical plasmas across a wide range of physical conditions, from molecular to fully ionized gas. It can handle number densities from as low as $\sim 10^{-8}~{\rm cm^{-3}}$ to as high as $\sim 10^{18}~{\rm cm^{-3}}$ and temperatures ranging from the cosmic microwave background at 2.72 K to $10^{10}$ K \citep{2013RMxAA..49..137F,2023RNAAS...7...45S}. Physical conditions such as the gas density and temperature of a TDE disk wind, fall within this range. CLOUDY models the optical spectral lines, He II (4686 \AA) and H$\alpha$ (6562 \AA) from TDE iPTF16fnl at three different epochs \citep{2017apj...844...46B} and also obtains the underlying physical conditions such as density, temperature and the element abundances. It requires a few input parameters: number density, chemical composition, radiation field, etc. Details about CLOUDY can be found in \citet{2013RMxAA..49..137F, 2017RMxAA..53..385F, 2022ApJ...934...53S} and references given there. All the models presented here are moderately optically thick to electron scattering, with $\tau_{\rm es} <3$ where $\tau_{\rm es}$ denotes the electron scattering optical depth.



Earlier, we modeled TDE emission lines from iPTF16axa using CLOUDY \citep{2023MNRAS.518.5693M}; We successfully reproduced the luminosities of the optical spectral lines, He II (4685.68 \AA) ~ and H$\alpha$ (H I 6562.80 \AA) for four epochs after the peak luminosity. Our model demonstrated that the enhanced He II / H$\alpha$ line ratio is due to the disruption of an evolved red giant star with super-solar He abundance and a smaller He II line optical depth. Note that the models and input parameters are similar to the iPTF16axa case, with one additional parameter, the filling factor, which accounts for the clumpiness of the medium.

%
\subsection{Models and input parameters}
\label{cloudy_parameters}
%

We adopt a similar assumption and set-up as those of \citet{2016ApJ...827....3R} and \citet{2023MNRAS.518.5693M} for our model. Specifically, we assume that a spherical atmosphere of ionized gas with the inner radius $r_l$ expands with a constant velocity of $v_{\rm w}$, and thus the size of the gas sphere gets larger proportionally to time. Unless otherwise noted in what follows, while the velocity is measured in SI units, the other quantities, such as the radius, density, luminosity, and so on, are measured in CGS units.
The observed FWHM of the He II and H$\alpha$ lines of iPTF16fnl reveal that an expanding velocities approximately equal 14000 km s$^{-1}$ and 10000 km s$^{-1}$, respectively  \citep{2017apj...844...46B}. We set $v_{\rm w}$ as a parameter to vary in a range close to the observed values. In all our models, we assume a number density follows $n_{l} (r/r_l)^{-2}$ \citep{{2016ApJ...827....3R},{2023MNRAS.518.5693M}}, where $n_{l}$ is the total hydrogen number density; $n_l=n_l$(H$^0$) + $n_l$(H$^+$) + 2$n_l$(H$_2$) +$\sum \limits_{i}$ $n_l$(H$_{i}$), where H$_{i}$ indicates the other species containing hydrogen nuclei such as H$_{3}^+$, H$_{2}^+$, etc. Hydrogen and helium make up more than $98\%$ of the ordinary matter in the universe. Hence, the mass density is calculated as 
\begin{eqnarray}
\rho_{l}
\simeq 
m_{\rm p}n_{l}
=1.67\times10^{-24}(1 + 4x)\,{\rm g~ cm^{-3}},
\label{eq:rho}
\end{eqnarray} 
where $m_{\rm p}$ is the proton mass, $x$ is the ratio of the helium number density to the total hydrogen number density $n_{l}$, and factor $4$ is coming from that the helium consists of two protons and two neutrons. Both $r_{l}$ and $n_{l}$ are set as free parameters, but $r_{l}$ remains the same for all three epochs, whereas the values of $n_{l}$ are different.

The previous works assumed that the atmosphere is not clumpy and spherically symmetric \citep{{2016ApJ...827....3R}, {2023MNRAS.518.5693M}}. 
However, a clumpy atmosphere has often been seen in astrophysical winds, e.g., Novae ejecta \citep{{2022ApJ...925..187P, 2019MNRAS.483.4884M}} and AGN winds \citep{2020MNRAS.498.4150D}. Moreover, \citet{2020MNRAS.494.4914P} showed that clumps lower the ionization state for disk winds in a TDE, suggesting that clumpiness is an important factor in deciding the physical states of the atmosphere. 
Therefore, we relax the assumption and introduce a simple parameter for clumpiness, the filling factor $f$. In our model, $f$ expresses a fraction of gas filling in a given volume, and the range is $0\le{f}\le1$. While $f=1$ indicates that the filling rate of a gas in the wind is $100\%$, $f=0$ means that the wind is in a vacuum. A certain middle value of $f$ shows that the wind consists of clumpy gas or the shape deviates from the spherically symmetric one we assumed for the wind geometry.
The $f$ parameter modifies the optical depth as 
\begin{eqnarray}
\tau(r)=
\int_{r_l}^{r}\, \alpha_{l,u} 
\biggr(
n_l - 
n_u 
\frac{g_l}{g_u}
\biggr) 
f 
\, dr
\label{eq:tau}
\end{eqnarray} 
as well as the volume emissivity 
\citep{2006hbic.book.....F}, where $\alpha_{l,u}$, $n_l$ $n_u$, $g_l$, and $g_u$ are line absorption cross section in ${\rm cm^2}$, lower level population, upper level population, lower level statistical weight, and upper level statistical weight, respectively.

As \citet{2020SSRv..216..114R} did so, we assume that the wind gas is irradiated by blackbody emission with a temperature $T_{l}$ at the wind's inner radius $r_{l}$ to produce the line emission. In our CLOUDY model, we treat $T_{l}$ as a free parameter. Given the density distribution, temperature $T_{l}$, and helium-to-hydrogen ratio $x$, CLOUDY simulates the wind's emission spectra from regions far beyond $r_{l}$, where the corresponding temperature is lower than $T_{l}$.


Since the host galaxy, Mrk950 has the solar metallicity \citep{2017apj...844...46B}, it is natural to adopt solar abundances as suggested by \citet{2010Ap&SS.328..179G}. In our model, we set the He abundance relative to H as a parameter so as to vary it close to its solar value, i.e., $\sim$ 0.1 
\citep{{1998SSRv...85..161G},{2010Ap&SS.328..179G}}. 
\citet{2017apj...844...46B} found the best fit for E(B-V) = 0 after considering the dereddening of the host galaxy. 
Moreover, the temperature is greater than the sublimation temperature of dust grains so that our model includes no dust.

Following our previous work on iPTF16axa \citep{2023MNRAS.518.5693M}, we adopt three different time-independent non-local Thermodynamic Equilibrium (NLTE) snapshot models for the three epochs ($<$30 days). One of them is an epoch earlier than the peak luminosity.

CLOUDY internally sets a permissible range of electron scattering optical depths. Hence, as well as our previous work \citep{2023MNRAS.518.5693M}, we adjust the model parameters such that the electron scattering optical depths remain within the permissible range. Note that all of our models are moderately optically thick to electron scattering.

%
\subsection{CLOUDY Results}
\label{cloudy_res}
%

We list the physical parameters at three epochs in Table~\ref{tab:table1}, which are calculated by applying the CLOUDY modeling to iPTF16fnl. Our CLOUDY modeling estimates $r_{l} = 10^{14.4}~{\rm cm}$ for all three epochs. We find that a temperature 10$^{4.9}$ K at $r_{l}$, due to incident radiation, is necessary to reproduce the observed line luminosities. \citet{2017apj...844...46B} found that the H$\alpha$ luminosity has a peak at day 0 among the three observational epochs -0.8, 0, and 29.3 days, and decreases with time. The CLOUDY-estimated line luminosity exhibits a similar behavior. Our models also predict the filling factor to be $ f = 0.8$. 

CLOUDY calculates gas temperature self-consistently from heating and cooling balance consisting of micro-physical processes \citep{2013RMxAA..49..137F}. Table~\ref{linelum} compares the observed and the model-predicted line luminosities with $x = 0.1$. The total heating and cooling powers for the three epochs are 10$^{41.353}$, 10$^{41.518}$, and 10$^{41.156}$ ergs s$^{-1}$, respectively. The electron temperature averaged over the thickness of the ionized gas for these three epochs are 1.80$\times$10$^4$, 1.70$\times$10$^4$, and 1.62$\times$10$^4$ K, respectively. The gas is fully ionized; Hydrogen and Helium are mainly in H$^+$, He$^{+}$, He$^{++}$.


\citet{2017apj...844...46B} have detected the He II 3203.08 \AA ~ line but did not provide the line luminosity. Our models predict strong He II 3203.08 \AA ~ lines and line luminosities of He II 3203.08 \AA ~ at these three epochs are, 1.8 $\times$10$^{39}$, 
2.7$\times$10$^{39}$, 1.1$\times$10$^{39}$ erg s$^{-1}$, respectively. While the model-predicted H$\alpha$ line luminosity is consistent with the observation, the predicted He II line luminosity is lower than the observed value.

\begin{figure}
\centering
\includegraphics[scale = 0.65]{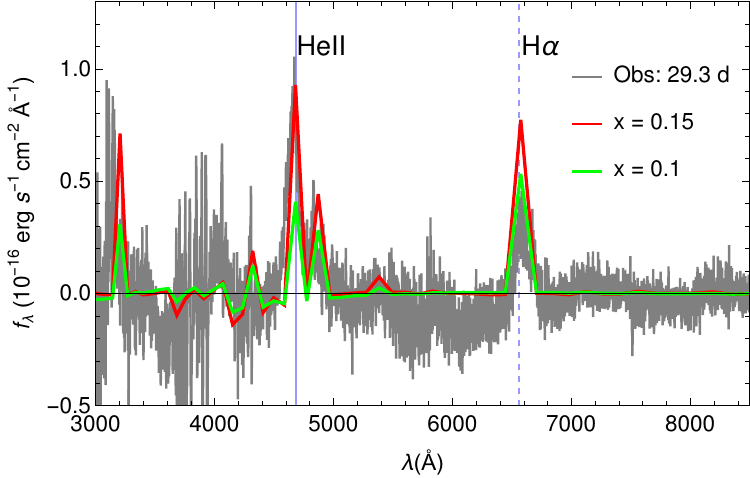}
\caption{
Comparison of the spectrum obtained from CLOUDY modeling with the observation for day 29.3. The gray line shows the observational data, while the green and red lines represent the CLOUDY models for helium-to-hydrogen ratios of $x = 0.1$ and $x = 0.15$, respectively. The vertical blue solid and dashed lines indicate the positions of the HeII (4686 \AA) and H$\alpha$ (6562 \AA) emission lines.
}
\label{fig:spec_cl}
\end{figure}

Moreover, we perform another set of modeling with $x = 0.15$ keeping other parameters the same. The results are listed in Table \ref{linelum1}. For the increased $x$, the total heating and total cooling powers at the three epochs are 10$^{41.443}$, 10$^{41.604}$, and 10$^{41.247}$ ergs s$^{-1}$, respectively. Similarly, the electron temperature averaged over the thickness of the ionized gas for these three epochs are 1.70$\times$10$^4$, 1.64$\times$10$^4$, 1.54$\times$10$^4$ K, respectively. 

Figure \ref{fig:spec_cl} compares the spectrum from the CLOUDY modeling for helium-to-hydrogen ratios of $x = 0.1$ and $x = 0.15$ with the observed spectrum from day 29.3. The He II line luminosity for $x = 0.15$ shows a better match with the observations compared to the case with $x = 0.1$.  Based on this spectroscopic analysis, we conclude that $x$ lies between 0.1 and 0.15. This range of $x$ implies that the disrupted star is likely to be a main sequence star because the solar helium to hydrogen number ratio is $\sim$ 0.1 (e.g., \citealt{{1998SSRv...85..161G},{2010Ap&SS.328..179G}}). 

\begin{table}
	\centering
	\caption{
Atmospheric parameter estimation by applying the CLOUDY to iPTF16fnl at three epochs. The different lines denote the different time epochs. The first, second, third, fourth, and fifth columns represent the time epoch, $r_{\rm l}$, $n_{\rm l}$, $T_{l}$, and $L$, respectively.
	}
	\label{tab:table1}
	\begin{tabular}{lllll} 
		\hline
	Days& $r_{l}$&$n_{l}$  & $T_{l}$ & $L$($T_{l}$) \\          
	&log (cm)&log (cm$^{-3}$) & log K & log (erg s$^{-1}$) \\ 
		\hline
		-0.8&14.4&  10.40&4.9 &42.9 \\
	    0&14.4& 10.45 & 4.9&43.0 \\
	    29.3&14.4& 10.20&4.9  &42.8 \\
	    \hline
		\end{tabular}
        \end{table}

\begin{table}
	\centering
	\caption{
Comparison between the observed and CLOUDY-modeled line luminosities at the three epochs. Each luminosity is shown in units of erg s$^{-1}$. The predicted line luminosity is estimated by the CLOUDY modeling with $x =0.1$. The first, second, and third columns display time epochs, HeII line luminosity, and H$\alpha$ line luminosity, respectively. 
	}
	\label{linelum}
 \scalebox{1}{
	\begin{tabular}{ccccc} 
		\hline
	Days&\multicolumn{2}{c}{He II (4685.68 \AA)} & \multicolumn{2}{c}{H$\alpha$ (6562.80 \AA)}\\     
	&&&&\\     
	&This work & Observed &This work & Observed\\ 
		\hline
		-0.8&  3.6$\times$10$^{39}$ & 1.5$\times$10$^{40}$$\pm$30\% & 5.2$\times$10$^{39}$& 3.8$\times$10$^{39}$$\pm$40\% \\
	    0& 5.4$\times$10$^{39}$ & 8.3$\times$10$^{39}$$\pm$30\% & 7.9$\times$10$^{39}$ & 6.6$\times$10$^{39}$$\pm$40\%\\
	    29.3& 2.2$\times$10$^{39}$ & 3.0$\times$10$^{39}$$\pm$30\%  &3.5$\times$10$^{39}$  & 3.0$\times$10$^{39}$$\pm$40\%\\
	    \hline
		\end{tabular}
  }
        \end{table}

\begin{table}
	\centering
	\caption{The same figure format but for $x = 0.15$.}
	\label{linelum1}
 \scalebox{1.0}{
	\begin{tabular}{ccccc} 
		\hline
	Days&\multicolumn{2}{c}{He II (4685.68 \AA)} & \multicolumn{2}{c}{H$\alpha$ (6562.80 \AA)}\\     
	&&&&\\      
	&This work & Observed &This work & Observed\\ 
		\hline
		-0.8&  6.5$\times$10$^{39}$ & 1.5$\times$10$^{40}$$\pm$30\% & 5.9$\times$10$^{39}$& 3.8$\times$10$^{39}$$\pm$40\%\\
	    0& 9.7$\times$10$^{39}$ & 8.3$\times$10$^{39}$$\pm$30\% & 9.1$\times$10$^{39}$ & 6.6$\times$10$^{39}$$\pm$40\%\\
	    29.3& 3.9$\times$10$^{39}$ & 3.0$\times$10$^{39}$$\pm$30\%  &4.1 $\times$10$^{39}$ & 3.0$\times$10$^{39}$$\pm$40\%\\
	    \hline
		\end{tabular}
  }
        \end{table}

The observations around day 0 show a full width at half maximum (FWHM) of $\sim 14000 \pm 3000\,{\rm km~s^{-1}}$ for HeII line and $\sim 10000 \pm 500~{\rm km~s^{-1}}$ for H$\alpha$ line. At late times, the FWHM reduces to $\sim 8500 \pm 1500\,{\rm km~s^{-1}}$ for HeII line and $6000 \pm 600~{\rm km~s^{-1}}$ for H$\alpha$ line \citep{2017apj...844...46B}. Using the CLOUDY-generated spectra, we estimate the FWHM by measuring the difference, $\Delta \lambda$, between points where the flux reaches half the line peak luminosity, and converting this to velocity using the relation $c\Delta\lambda / \lambda$. The CLOUDY model predicts an FWHM of $\sim 6500$ km s$^{-1}$ for the HeII line and $\sim 6000$ km s$^{-1}$ for the H$\alpha$ line at day 29.3. While the FWHM of HeII is underestimated at late times, the H$\alpha$ line agrees with the observations.

\begin{figure}
\centering
\includegraphics[scale = 0.6]{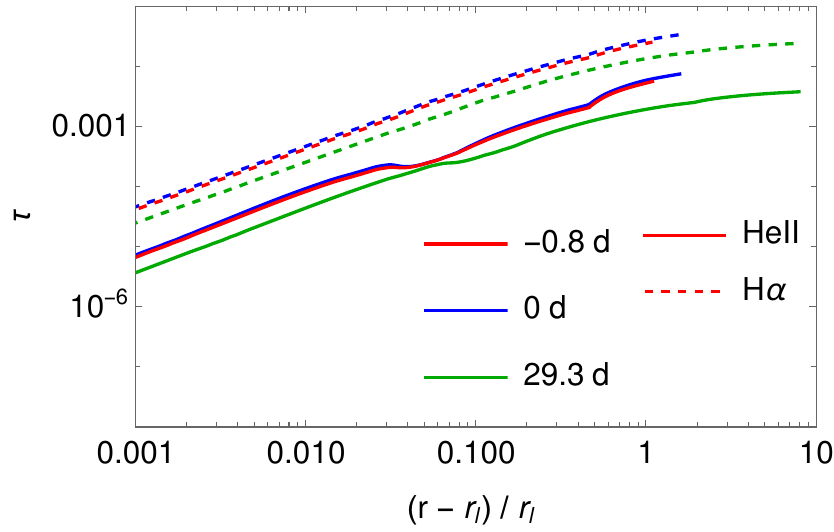}
\caption{
Comparison of the optical-depth radial profiles between the Helium and Hydrogen emission lines obtained by the CLOUDY modeling. The relative radius, $r-r_l$, is normalized by the wind inner radius, $r_l$. The different colors denote the different time epochs. The solid and dashed lines represent the optical depth of the helium (HeII) and hydrogen (H$\alpha$) emission lines, respectively.
}
\label{fig:tau}
\end{figure}

Figure~\ref{fig:tau} depicts the radial profile of the optical depths of the Helium and Hydrogen emission lines. Those optical depths are calculated by equation~(\ref{eq:tau}). It is noted from the figure that the Helium emission line's optical depth is about two orders of magnitude lower than that of the Hydrogen lines at all three observational times. The scattering and self-absorption of photons increase with optical depth, which depends on the medium's density, temperature, and photon frequency. The line luminosity decreases as $L \propto {\rm e}^{-\tau}B(\nu,~T)$, where $\tau$ is the optical depth corresponding to the emission line and $B(\nu,~T)$ is the Planck blackbody function with temperature $T$. Thus, the lower optical depth for the helium emission line makes the helium line luminosity higher than the hydrogen line luminosity. This result suggests that the observed dominance of the helium line of iPTF16fnl arises from the optical depth effects.

Let us describe finally that our current models have some limitations.
Firstly, the geometry is unknown and might be unknowable. Secondly, the clumpiness may be more complex than our simplistic approach. At earlier epochs, time-dependent CLOUDY modeling would do better but it is significantly slower than the time-independent CLOUDY modeling. This is because time-dependent models account for physical processes such as ionization/recombination and cooling/heating rates, which evolve over time and require tracking numerous intermediate steps, thereby substantially increasing the computational time. Finally, we vary model parameters such that the electron scattering optical depths remain within the default permissible range of CLOUDY. Note that these can affect the results that described in this section.

%
%

%
\section{Disk-wind model}
\label{dwfit}
%

This section briefly describes the disk-wind model (see the appendix \ref{dwmodel_theory} for the details) for fitting the observed continuum spectrum to get values of physical parameters of iPTF16fnl. This model assumes a steady-state, slim disk with an outflow that constitutes the atmosphere \citep{2023MNRAS.518.5693M}.
Because the mass accretion rate exceeds the Eddington rate in the slim disk model, the intense radiation pressure induces an outflow \citep{2009MNRAS.400.2070S,2015mnras.448.3514c,2019ApJ...885...93F}.  The mass conservation law results in that the sum of the mass accretion rate ($\dot{M}_{\rm a}$) and mass outflow rate ($\dot{M}_{\rm out}$) equals the mass fallback rate of the disrupted stellar debris ($\dot{M}_{\rm fb}$), i.e., $\dot{M}_{\rm a} + \dot{M}_{\rm out} = \dot{M}_{\rm fb}$, 
where 
\begin{equation}
\dot{M}_{\rm fb} = \frac{1}{3}\frac{M_{\star}}{t_{\rm m}} \left(\frac{t_{\rm m} + t}{t_{\rm m}}\right)^{-5/3},
\label{mfb}
\end{equation}
and the period of the most tightly bound debris is given by
\begin{eqnarray}
 t_{\rm m} = 40.8~{\rm days}\,
\left(
\frac{M_{\bullet}}{10^6 M_{\odot}}
\right)^{1/2} 
 \left(
 \frac{M_{\star}}{M_\odot} 
 \right)^{1/5}
 k^{-3/2}
 \label{eq:tm}
\end{eqnarray} 
with the radius of star $R_{\star}=R_{\odot} (M_{\star}/M_{\odot})^{0.8}$ \citep{1994sse..book.....K}. Here, $k$ is the tidal spin-up factor to take into account the spin-up of a star due to the tidal torque by the black hole \citep{2001ApJ...549..948A,2020MNRAS.496.1784M}. The $k = 1$ means the tidal torque is neglected, whereas $k = 3$ indicates that the tidal torque spins up the star to its maximum rotational velocity where the centrifugal force exceeds the stellar self-gravity leading to stellar disruption \citep{2002ApJ...576..753L}. 
In our calculation, the time origin $t = 0$ corresponds to the fallback time of the most tightly bound debris.



We also assume the mass accretion rate and mass outflow rate are 
\begin{eqnarray}
\dot{M}_{\rm out}&=&f_0\,\dot{M}_{\rm fb}, \label{eq:mdotout} \\
\dot{M}_{\rm a}&=&(1-f_0)\dot{M}_{\rm fb}, 
\label{eq:mdotacc}
\end{eqnarray}
respectively, where $f_0$ is given by \citep{2015ApJ...814..141M}
\begin{equation}
f_0 (\eta) = {\rm Max}\left[\frac{2}{\pi} {\rm Tan^{-1}}\left\{\frac{1}{4.5}\left(\eta\frac{\dot{M}_{\rm fb}c^2}{L_{\rm Edd}}-1\right)\right\},~0\right]
\label{f0eqt}
\end{equation}
and $\eta$ is the radiative efficiency:
\begin{eqnarray}
\eta 
&=& 
\frac{L_{\rm b}}{\dot{M}_{\rm a}c^2} 
\nonumber \\
&=&
\frac{3R_{\rm S}}{2}  \int_{R_{\rm in}}^{R_{\rm out}} \frac{f_R}{R^2} \left[\frac{1}{2} + \right. \nonumber \\ 
&& \left. \sqrt{\frac{1}{4}+ \frac{3}{2} f_R \left(\frac{R_{\rm S}}{R}\right)^2 \left(\frac{\dot{M}_{\rm a}}{L_{\rm Edd}/c^2}\right)^2}~\right]^{-1} \, \diff R,
\label{etadisk}
\end{eqnarray}
where we use equations~(\ref{tefdisc}) and (\ref{lumb}) for the derivation, $R_{\rm S} = 2 G M_{\bullet}/c^2$ is the Schwarzschild radius, $f_R = 1 - \sqrt{R_{\rm in} / R}$, $R_{\rm in}$ is the disk inner radius taken to be innermost stable circular orbit of non-spinning black hole, i.e. $R_{\rm in} = 3 R_{\rm S}$. Note that $\eta$ is not constant for the super-Eddington phase and varies with mass accretion rate. 

The disk-wind model, detailed in Appendix \ref{dwmodel_theory}, assumes a spherical outflow characterized by a constant wind inner radius, $r_l$, and wind velocity, $v_{\rm w}$, leading to a radial density profile of $\rho(r) \propto r^{-2}$. The parameter $r$ represents the radial distance in the outflow, while $R$ denotes the disk's radial coordinate. We assume the outer radius $r_{\rm w,0}$ is much larger than $r_l$, allowing for the calculation of model quantities without specifying the outer radius. The wind's photospheric radius, $r_{\rm ph}$, and temperature, $T_{\rm ph}$, are described by equations (\ref{eq:rph}) and (\ref{eq:tph}), respectively, depends on $\dot{M}_{\rm out}$, $r_l$, and $v_{\rm w}$. Using equations (\ref{mfb}), (\ref{eq:mdotout}), and (\ref{f0eqt}), that gives $\dot{M}_{\rm out}$, the wind spectral luminosity given by equation (\ref{lspec}), is a function of $M_{\bullet}$, $M_{\star}$, $r_{l}$, and $v_{\rm w}$.

There are several unknown parameters: black hole mass $M_{\bullet}$, stellar mass $M_{\star}$, wind's inner radius $r_l$, and velocity $v_{\rm w}$. The stellar debris circularizes to form an accretion disk, but the circularization time for the debris is unknown yet very much. Therefore, we also introduce a time parameter $\Delta t$ in the mass fallback rate, which delineates the shift in time $t$ to describe the ambiguity in the starting time of disk accretion after the innermost debris returns to the pericenter. Fitting the disk-wind model to the observed spectral continuum makes it possible to estimate the values of these five unknown parameters directly. The other quantities, such as density, temperatures, and radius of the photosphere, are obtained with these parameters. We will do them in the next section.

%
\subsection{Estimation of five parameters and disk-wind structure}
\label{dwres}
%

This section estimates the five basic physical quantities and parameters by comparing the observed photometry data and the disk-wind model using the likelihood analysis. Subsequently, those values are used to explore the time evolution of the disk and wind emissions. First, we obtain the photometry data of iPTF16fnl from \url{https://cdsarc.cds.unistra.fr/viz-bin/cat/J/ApJ/844/46#/browse}. The host galaxy contaminates the Swift photometry and has very little flux in the Swift UV bands, but the flux dominates in the U, B, and V bands \citep{2017apj...844...46B}. In our calculation, we neglect Swift U, B, and V bands and instead use the Swift UVW1, UVM2, and UVW2 bands along with the u, g, r, and i bands. To get a continuum with statistically sufficient points, we need to select such observations in different bands that each observed time is nearly equal; at least, the difference in an observed time width of the respective epochs is within a day. In our case, we successfully collected 12 observational epochs, where the difference in the observed time width is $0.7$ days at maximum.

%
%
\begin{figure*}
\centering
\includegraphics[scale = 0.62]{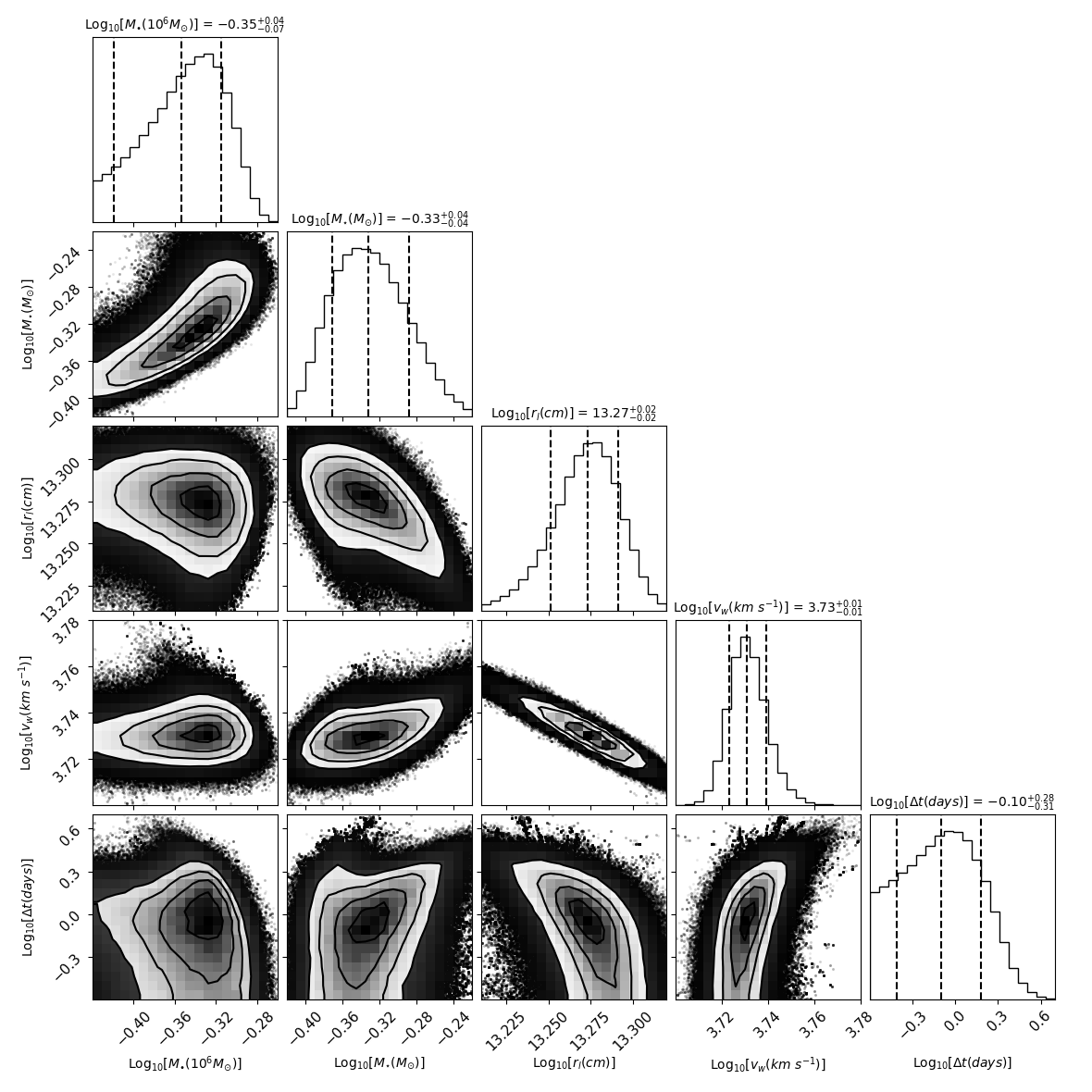}
\caption{Corner plot of the disk-wind model fit to the continuum, which contains 10000 MCMC simulations. See section \ref{dwres}.
}
\label{mbhcnt}
\end{figure*}

The likelihood at an epoch $i$ is given by $\mathcal{L}_{i} = (\prod_{n = 1}^{n = N} \sqrt{2 \pi} \sigma_{\rm i,n})^{-1} \exp(-\chi_i^2 / 2)$, where $N$ is total number observational points in the epoch $i$, $\sigma_{\rm i,n}$ is the observed uncertainty of the $n$th observation in $i$th epoch, and $\chi_i^2$ is given by $\chi_i^2 = \sum_{n = 1}^{n = N} \left[L^{\rm w}(\nu_{\rm i,n}) - L_{\rm i,n}{\rm (obs)}\right]^2 / \sigma_{\rm i,n}^2$. The $L^{\rm w}(\nu_{\rm i,n})$ is the spectral luminosity given by equation (\ref{lspec}) for frequency $\nu$ corresponding to $n$th observation and $L_{\rm i,n}{\rm (obs)}$ is the observed luminosity. Since we have 12 epochs of observations, the total likelihood is $\mathcal{L} = \prod_{i = 1}^{i = 12} \mathcal{L}_i$, which results in total $\chi^2$ given by
\begin{equation}
\chi^2 = \sum_{i =1}^{i = 12} \chi_i^2 = \sum_{i =1}^{i = 12} \sum_{n = 1}^{n = N} \frac{\left[L^{\rm w}(\nu_{\rm i,n}) - L_{\rm i,n}{\rm (obs)}\right]^2}{\sigma_{\rm i,n}^2}.
\end{equation}
We perform a $\chi^2$ minimization simultaneously on all observational data points to obtain the best-fitting parameters: black hole mass, stellar mass, wind inner radius, and wind velocity. We perform Markov chain Monte Carlo (MCMC) simulations using the Python-based emcee code \citep{2013PASP..125..306F} to estimate the physical parameters and their standard errors, as well as to generate the corner plot.

%
%
\begin{figure*}
\centering
\includegraphics[scale = 0.45]{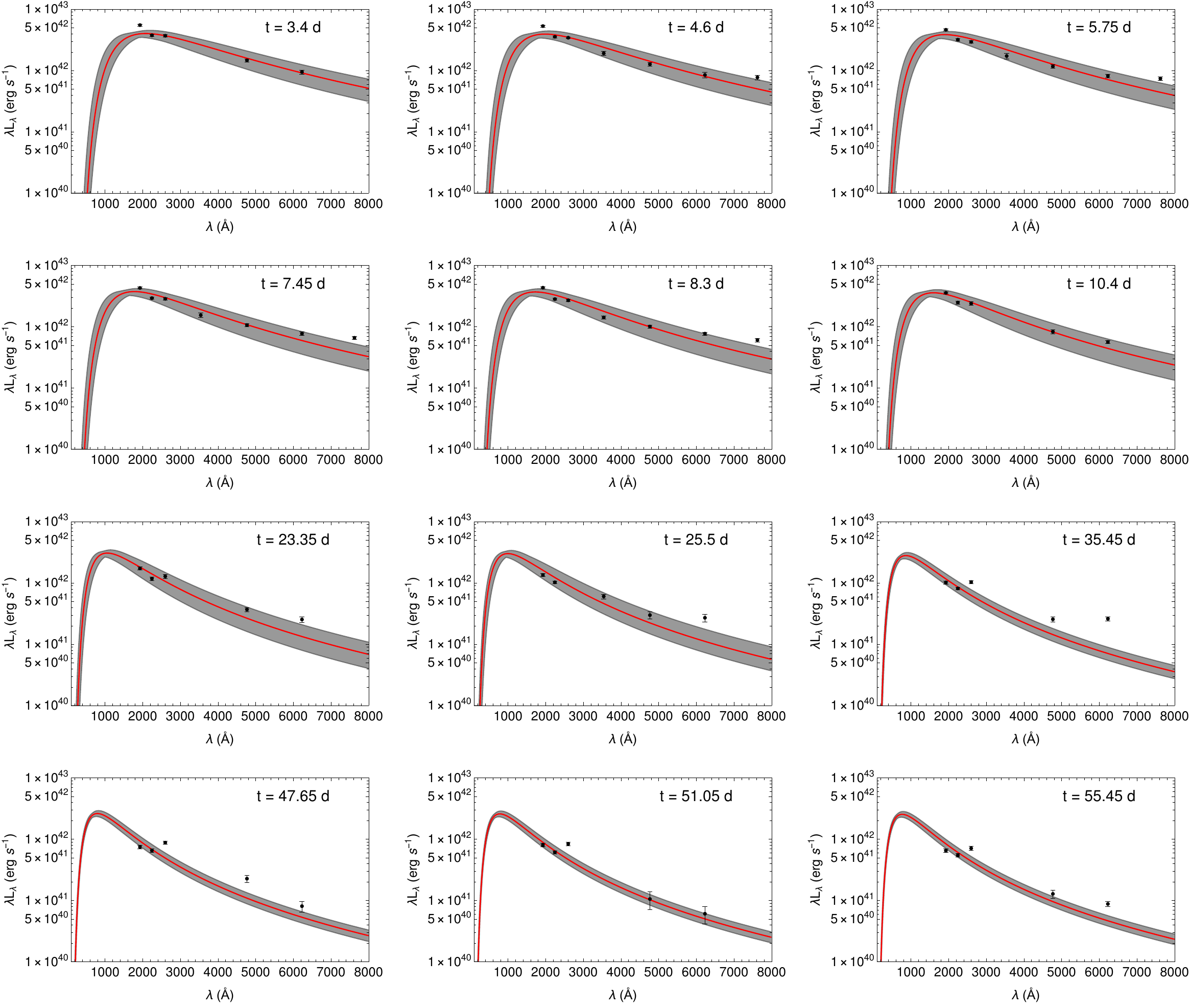}
\caption{
Continuum spectra of iPTF16fnl. The black-filled circles with error bars and the red solid line indicate the observational data and theoretical spectra with the five parameters (see equation~\ref{lspec} for the detail), respectively. The gray-shaded region represents the error range of the theoretical spectrum caused by the five parameters. The corresponding observational epochs are shown at the top right of each panel. See section \ref{dwres}.
}
\label{mbhlspecplt}
\end{figure*}

The resultant five parameters are summarized as follows: $\displaystyle{M_{\bullet} = 4.43^{+0.17}_{-0.14} \times 10^5 M_{\odot}}$, $\displaystyle{M_{\star} = 0.46^{+0.12}_{-0.06} M_{\odot}}$, $\displaystyle{r_l = 1.88^{+0.293}_{-0.224} \times 10^{13}~{\rm cm}}$, $\displaystyle{v_{\rm w}  = 5382.7^{+359.4}_{-347.1} ~{\rm km~s^{-1}}}$, and $\displaystyle{\Delta t = 0.79^{+1.24}_{-0.572} ~{\rm days}}$. Note that the wind velocity is close to the full-width half maximum (FWHM) of the H$\alpha$ ($ 6000 \pm 600~{\rm km~s^{-1}}$) lines at the late time of the observations (see Figure 9 in \citealp{2017apj...844...46B}).
The wind's inner radius, in terms of the tidal radius, is $r_{l} = 5.05^{+2.7}_{-1.1}\, r_{\rm t}$, which is larger than the circularization radius (typical TDE disk size) of $2 r_{\rm t}$ \citep{1999ApJ...514..180U,2009MNRAS.400.2070S,2016MNRAS.461.3760H}.

Figure~\ref{mbhcnt} shows the corner plot from our model fit to the continuum using MCMC simulations. The elongated shapes of the contours, rather than a circular one, indicates parameter correlations. For example, increasing $M_{\star}$ or decreasing $M_{\bullet}$ leads to a higher mass fallback rate, which enhances the mass outflow rate and boosts the wind emission. On the other hand, increasing $r_{l}$ reduces the wind density and temperature, thereby decreasing the wind emission. This interplay among the parameters creates the observed degeneracies in the corner plot. Figure~\ref{mbhlspecplt} makes a comparison between the theoretical spectra with the five parameters and the observed spectra on respective epochs. It is noted from the figure that the theoretical spectra agree well with the observed ones.

%
%
\begin{figure}
\centering
\includegraphics[scale = 0.65]{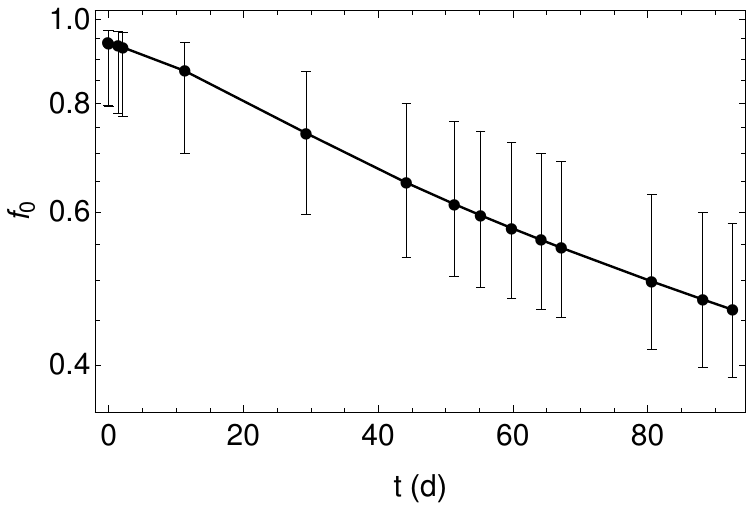}
\includegraphics[scale = 0.65]{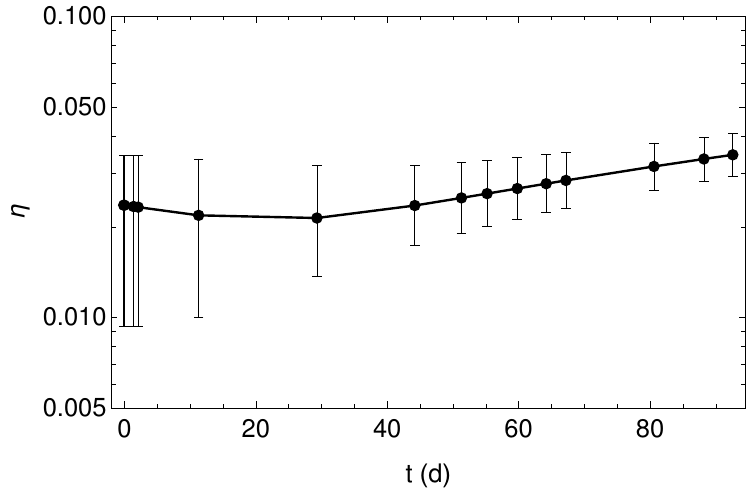}
\caption{
Time evolution of the fraction of mass outflow to fallback rates ($f_0$)  and the radiative efficiency ($\eta$) connecting between the bolometric luminosity and mass accretion rate. The upper and lower panels display the time evolution of $f_0$ and $\eta$, respectively. In both panels, while the black circles indicate $f_0$ or $\eta$, which are estimated by equations (\ref{f0eqt}) and (\ref{etadisk}) with the mean value of the five parameters, the vertical error bars represent the error range caused by the standard deviation of the five parameters.
}
\label{etaplt}
\end{figure}

%
%
\begin{figure}
\centering
\includegraphics[scale = 0.53]{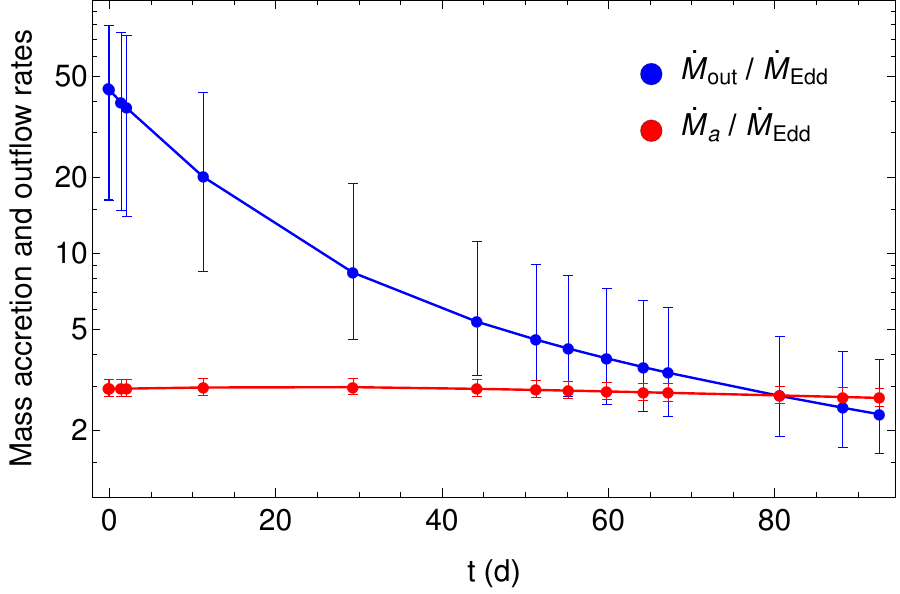}
\caption{
Time dependence of the wind mass outflow rate and disk accretion rate. Both rates are normalized by the Eddington accretion rate.
The blue and red lines connected between each filled circle with lines display the mass outflow and accretion rates, respectively.
The filled circles delineate the mean value of the five parameters, whereas the error bar per circle corresponds to the standard deviation of the five parameters.
}
\label{moutplt}
\end{figure}

According to equations (\ref{f0eqt}) and (\ref{etadisk}) with equation (\ref{eq:mdotacc}), $f_0$ and $\eta$ are a function of time and a complicated dependence on each other. Two panels of Figure~\ref{etaplt} depict the solutions for $f_0$ and $\eta$, respectively. It is noted from the figure that $f_0$ decreases with time, while $\eta$ is almost constant to be a value between $0.01$ and $0.04$ over the observational time.

Figure~\ref{moutplt} depicts the wind mass outflow rate and the mass accretion rate of the disk. 
The Eddington accretion rate,
\begin{equation}
\dot{M}_{\rm Edd} = \frac{1}{\eta}\frac{L_{\rm Edd}}{c^2},
\label{Medd}
\end{equation}
normalizes both rates. It is noted from the figure that the mass outflow rate is higher than the mass accretion rate and also decreases with time because of $\dot{M}_{\rm out}=f_0\dot{M}_{\rm fb}$. Since $f_0$ decreases as the mass fallback rate declines (see equation~\ref{mfb}), the time evolution of the mass outflow rate deviates from the $t^{-5/3}$ law of the mass fallback rate. 
In contrast, the figure shows that the mass accretion rate is almost constant to be $\dot{M}_{\rm a} \approx 2.9 \dot{M}_{\rm Edd}$ over time. This is because the increment in $1 -f_0$ is comparable to the decline in $\dot{M}_{\rm fb}$ for a given mass accretion rate $\dot{M}_{\rm a} = (1 - f_0)\dot{M}_{\rm fb}$, in other words, the effects of those two terms cancel each other out. The constancy of the mass accretion rate is consistent with the steady-state slim disk model.

%
\begin{figure}
\centering
\subfigure[]{\includegraphics[scale = 0.65]{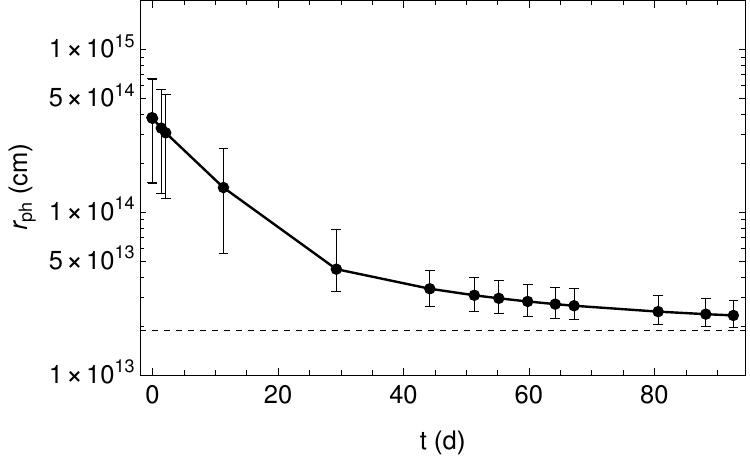}}
\subfigure[]{\includegraphics[scale = 0.65]{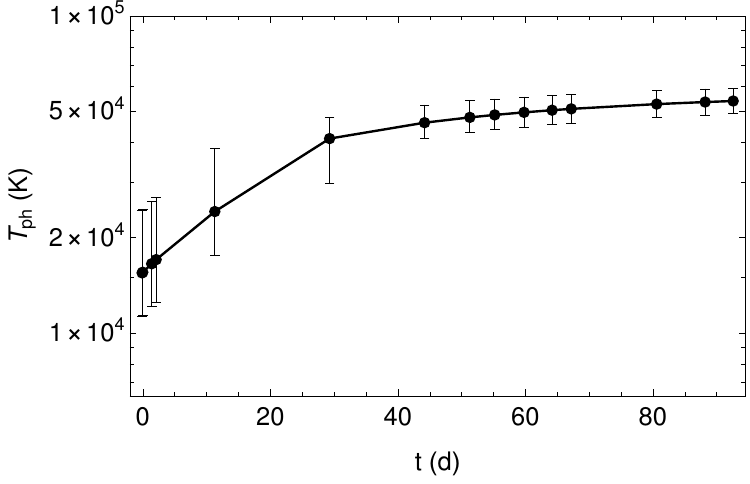}}
\caption{
Time evolution of the photospheric radius and the corresponding thermal temperature. Both $r_{\rm ph}$ and $T_{\rm ph}$ are estimated by equation~(\ref{eq:rph}) and (\ref{eq:tph}). Panel (a) denotes the time dependence of $r_{\rm ph}$, whereas panel (b) depicts that of $T_{\rm ph}$. In panel (a), the black dashed line represents the wind's inner radius $r_{l}$. In both panels, the black circles indicate the value estimated using the mean value of the five parameters. The vertical lines represent the error range caused by the standard deviation of the five parameters. 
}
\label{mbhrphTph}
\end{figure}

Figure~\ref{mbhrphTph} demonstrates the photospheric radius and temperature of the wind estimated using equations (\ref{eq:rph}) and (\ref{eq:tph}), respectively. It is not trivial to see their time dependency from those equations. However, considering the outflow is expanding with $\dot{M}_{\rm out}\propto t^{-5/3}$ at the constant velocity, the wind density decreases with time, resulting in the photospheric radius is predicted to be smaller with time. Note that the range of the photospheric radius agrees with the blackbody radius ($\sim (0.3 - 2)\times 10^{14}~{\rm cm}$) obtained by Blagorodonova et al. (2017). In contrast, the photosphere temperature increases with time and lies in the range $T_{\rm ph} \in \{1.5,~5\} \times 10^4 ~{\rm K}$. The increase in the photosphere temperature is slight at late times. The blackbody temperature of the photosphere that \citet{2017apj...844...46B} provided is $\sim 2 \times 10^4~{\rm K}$. This is also in good agreement with our photospheric temperature.

%
%
\begin{figure}
\centering
\includegraphics[scale = 0.67]{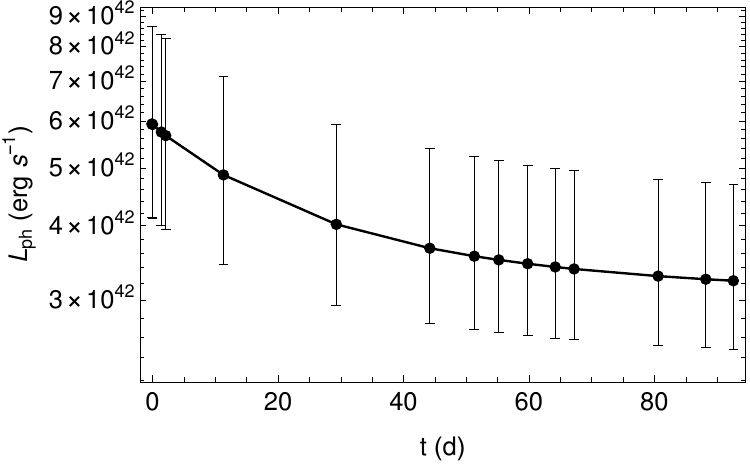}
\caption{
Light curve of the luminosity at the trapping radius (see equation~\ref{lwind}). The black-filled circles indicate the luminosity estimated using the mean value of the five parameters. The vertical line per circle represents the error bar corresponding to the standard deviation of the five parameters.
}
\label{mbhlwind}
\end{figure}

Figure~\ref{mbhlwind} shows the light variation of the wind luminosity given by equation (\ref{lwind}) in the diffusion regime of the wind. It is noted that the wind luminosity decreases with time. The peak luminosity is $\sim 6^{+3}_{-2} \times 10^{42}~{\rm erg~s^{-1}}$, which, within its error limits, is close to the peak bolometric luminosity of $\sim 10^{43}~{\rm erg~s^{-1}}$ estimated by Blagorodonova et al. (2017) based on a single temperature blackbody model.

%
%
\begin{table}
\caption{
The disk bolometric luminosity and spectral luminosity of the X-ray bands (0.3-10\,keV) at ten observational times. The super and subscripts indicate the upper and lower limits of the error range, which corresponds to the standard deviation of the five parameters. The first, second, and third columns represent the observational time, disk bolometric luminosity ($L_{\rm b}$), and spectral X-ray luminosity ($L_{\rm X}$), respectively. 
}

\label{mbhlbolt}
\centering
\begin{tabular}{|c|c|c|}
\hline
&&\\
$t~{\rm (days)}$ & $L_{\rm b}~{\rm (10^{44}~ erg~s^{-1})}$ & $L_{\rm X}~{\rm (10^{40}~ erg~s^{-1})}$\\
&&\\
\hline 
&&\\
1.4 & $1.91^{+0.29}_{-0.25}$ & $2.06^{+0.09}_{-0.13}$ \\
&&\\
11.3 & $1.93^{+0.26}_{-0.26}$ & $2.08^{+0.09}_{-0.17}$ \\
&&\\
29.3 & $1.94^{+0.26}_{-0.25}$ & $2.08^{+0.10}_{-0.23}$ \\
&&\\
44.2 & $1.91^{+0.26}_{-0.24}$ & $2.06^{+0.12}_{-0.26}$ \\
&&\\
51.3 & $1.89^{+0.27}_{-0.24}$ & $2.05^{+0.13}_{-0.27}$ \\
&&\\
55.2 & $1.88^{+0.27}_{-0.24}$ & $2.04^{+0.14}_{-0.28}$ \\
&&\\
88.2 & $1.77^{+0.26}_{-0.23}$ & $1.96^{+0.19}_{-0.32}$ \\
&&\\
\hline
\end{tabular}
\end{table}

%
%
\begin{figure}
\centering
\includegraphics[scale = 0.63]{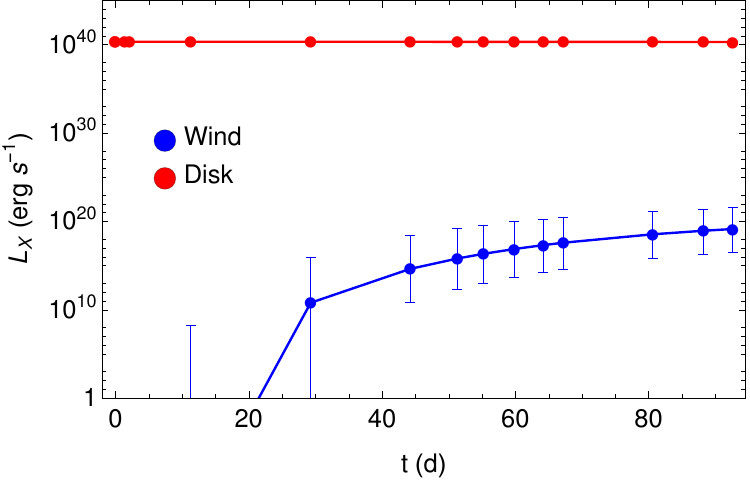}
\caption{
Light curves of X-ray emissions from the disk and wind. The wind X-ray luminosity is calculated at the photospheric radius (see equation~\ref{eq:lx_rph}), while the disk X-ray luminosity is estimated by equation~(\ref{xlum}). The red- and blue-filled circles indicate the disk and wind X-ray luminosities, respectively, calculated using the five parameters' mean values. The vertical line per circle represents the error bar corresponding to the standard deviation of the five parameters. Note that the error bars of the disk X-ray luminosity are too small to see in the plot; instead, see the error range in Table~\ref{mbhlbolt}.
}
\label{fig:lx}
\end{figure}

Table~\ref{mbhlbolt} shows the disk bolometric and X-ray luminosities, which are estimated by equations (\ref{lumb}) and (\ref{xlum}), respectively. The disk bolometric luminosity, $L_{\rm b} \sim 1.9 \times 10^{44}~{\rm erg~s^{-1}}$, exceeds the Eddington luminosity, $L_{\rm Edd} = 6.5^{+2.5}_{-2.1} \times 10^{43}~{\rm erg~s^{-1}}$, for the derived black hole mass by a factor of $\sim 2.9$. This suggests most of the photons emitted from the disk are scattered and absorbed in the media of the outflowing wind so that the detectable photons are emitted at the photosphere of the wind. This results in that the disk bolometric luminosity is reduced to $\lesssim 10^{43}~{\rm erg~s^{-1}}$ by more than an order of magnitude and observed in the OUV wave bands. In contrast, the X-ray photons are radiated from the sufficiently inner part of the disk. In fact, using equation (\ref{tefdisc}), the peak value of the effective disk temperature is $\sim 2.3 \times 10^5~{\rm K}$, which is about one order of magnitude higher than the photosphere temperature in the wind. As seen in Table~\ref{mbhlbolt}, the disk X-ray luminosity is estimated to be $L_{\rm X}\sim 10^{40}~{\rm erg~s^{-1}}$. Figure~\ref{fig:lx} compares the X-ray light curves of disk and wind emissions, where the wind X-ray emission originates from the photosphere of the wind. It is clear from the figure that the disk X-ray luminosity is much higher than the wind X-ray luminosity, which is calculated by equation~(\ref{eq:lx_rph}). The disk X-ray luminosity, $\sim 2 \times 10^{40}~{\rm erg~s^{-1}}$, is an order of magnitude higher than the observed X-ray luminosity, $L_{\rm X} \sim 2.4^{+1.9}_{-1.1} \times 10^{39}~{\rm erg~s^{-1}}$, that Blagorodonova et al. (2017) estimated by stacking all the XRT data over the observation period together (58 ks of total exposure time).

Finally, let us compare the wind densities between CLOUDY and disk-wind models. As seen in Tables~\ref{tab:table1} and~\ref{tbl:mbhdencomp}, the inner radius estimated from CLOUDY modeling is larger than $r_{\rm l}$, which is calculated from the disk-wind model. Both models assume the mass density to be $\rho \propto r^{-2}$. We compare the wind density estimated from disk-wind and CLOUDY models at the inner radius obtained by CLOUDY.  From equation~\ref{eq:rho}, the number density depends on $x$, which is the number density ratio of He over H. 
Table~\ref{tbl:mbhdencomp} displays the wind number densities estimated for $x =0.1$ and $x=0.15$, along with the electron scattering optical depth $\tau_{\rm es}$, calculated using equation (\ref{eq:taues}) from the disk-wind model. The number densities from the disk-wind model show good agreement with those from the CLOUDY model within the error limits.
However, $\tau_{\rm es}$ significantly exceeds the CLOUDY range of $\tau_{\rm es} < 5$ in the earlier two epochs. According to equation~(\ref{eq:taues}), $\tau_{\rm es}$ becomes small enough to be in the CLOUDY region at the later time when $\dot{M}\lesssim\dot{M}_{\rm Edd}$. This result indicates that our disk-wind model is applicable to the CLOUDY model only in the late epoch.

%
%
\begin{table}
\caption{
Comparison of the wind number density estimated at $r_{\rm l} = 2.51 \times 10^{14}~{\rm cm}$ between disk-wind model and CLOUDY modeling. The table also includes the electron scattering optical depth estimated using equation (\ref{eq:taues}). The first, second, third, and fourth columns represent the observational time, CLOUDY number density, disk-wind model number density, and the optical depth respectively. For the number density of the disk-wind model, two values of $x$ are adopted.} 
\label{tbl:mbhdencomp}
 \centering
\begin{tabular}{ccccc}
\hline
& CLOUDY & \multicolumn{3}{c}{Disk-wind}   \\
\hline
&&&&\\
$t~{\rm (days)}$ & $\log_{\rm 10}(n_l~{\rm (cm^{-3})})$ & \multicolumn{2}{c}{$\log_{\rm 10}(n_l~{\rm (cm^{-3})})$} & $\tau_{\rm es}$ \\
&&&&\\
&& ($x = 0.1$) & ($x = 0.15$) & \\
\hline 
&&&&\\
-0.8 & 10.40 & 11.14$^{+0.34}_{-0.61}$ & 11.08$^{+0.34}_{-0.61}$ & 27.5$^{+15.3}_{-19.1}$\\
&&&&\\
0.0 & 10.45 & 11.13$^{+0.34}_{-0.62}$ & 11.07$^{+0.34}_{-0.61}$ & 27.1$^{+15.1}_{-18.9}$\\
&&&&\\
29.3 & 10.20 & 10.44$^{+0.24}_{-0.27}$ & 10.38$^{+0.24}_{-0.28}$ & 4.96$^{+2.73}_{-3.1}$\\
&&&&\\
\hline
\end{tabular}
\end{table}

%
\section{Discussion}
\label{discuss}
%

iPTF16fnl is an optically discovered low luminosity TDE, and the emission dominates in the UV/optical bands without observationally significant X-ray emission. The host of iPTF16fnl is a post-starburst galaxy with an observed host velocity dispersion of $89 \pm 1~{\rm km~s^{-1}}$ \citep{2017apj...844...46B}, which were estimated from the CaII $\lambda \lambda$8544,8664 absorption lines. While the black hole mass inferred from the $M_{\bullet}-\sigma$ relation \citep{2013ApJ...764..184M} is $(2.138 \pm 1.87) \times 10^6 M_{\odot}$ \citep{2017apj...844...46B}, our model provides the black hole mass to be $\displaystyle{M_{\bullet} = 4.43^{+0.17}_{-0.14} \times 10^5 M_{\odot}}$, which agrees with the previous estimate within its error limit. The host galaxy has a metallicity of $Z = 0.18$, which nearly equals the solar metallicity $Z_{\odot} = 0.2$ \citep{2017apj...844...46B}, suggesting that the host galaxy should have more young main-sequence stars. The elemental abundances in the atmosphere that our CLOUDY modeling provides support this speculation. In that way, our method is useful in probing the nature of the disrupted star. 

Our disk-wind model have obtained the stellar mass of $M_{\star} = 0.46^{+0.12}_{-0.06} M_{\odot}$. In contrast, the MosFIT model applied to iPTF16fnl provides a smaller stellar mass of $M_{\star} = 0.101^{+0.008}_{-0.004} M_{\odot}$ \citep{2019ApJ...872..151M}. The MosFIT model considers no detailed geometry and radiative processes of the wind. Moreover, it assumes a constant radiative efficiency and models the photospheric radius as a power-law function of the mass fallback rate, where the power-law index is a constant parameter determined through model fitting to observations. These assumptions give a flexibility to model the photospheric temperature evolution and emission without considering in details the geometry and radiative processes in the atmosphere. On the other hand, the CLOUDY modeling demonstrates that the wind is clumpy with a filling factor of 0.8, for the $f = 1$ assumption results in more discrepancy in the He II $\lambda$ 4686 line. In addition, from our disk-wind model, the photospheric radius and temperature depend on the wind's inner radius, mass fallback rate, and radiative efficiency, which is not constant (see equations~\ref{etadisk}, \ref{eq:rph}, and \ref{eq:tph}). 
The deviation of the stellar mass between ours and the MosFit model \citep{2019ApJ...872..151M} demonstrates the importance of considering the detailed wind geometry and radiative processes for stellar mass estimation.

\citet{2016MNRAS.458..127K} used Modules for Experiments in Stellar Astrophysics (MESA) model to study the time evolution of the element abundances within the star, assuming the star has the solar metallicity at the initial time. Since the Sun's helium and other elements fractions are significantly smaller than hydrogen, the helium-to-hydrogen number ratio is set to be $\sim$ 0.09 at the initial time. For a $0.5\,M_{\odot}$ star, he showed that the abundances of helium, carbon, and nitrogen do not change appreciably over a timescale of 10 Gyr, indicating that low-mass stars do not have sufficient time to increase their helium abundance considering the age of the universe.
This complements our CLOUDY estimation of a small helium-to-hydrogen number ratio, indicating that the disrupted star has a low helium abundance. In this work, we have considered the complete disruption of a star and estimated the helium/hydrogen number ratio to be between 0.1 and 0.15, which suggests that the star is likely a main sequence star. However, a partial disruption of even a metal rich star (e.g., red giant stars) could also explain the low helium/hydrogen number ratio, as the outer shell of a star is primarily composed of hydrogen.



In our disk-wind model, the radiative efficiency $\eta$ varies with time due to the time-dependent mass fallback rate, as shown in equation (\ref{etadisk}). The proportionality coefficient $f_0$ goes to zero if $\dot{M}_{\rm fb} \leq \dot{M}_{\rm Edd}$, indicating no outflow so that the radiative efficiency is constant in time because $f_0$ is independent of the mass fallback rate. Moreover, our calculation demonstrates that $0.01\lesssim\eta\lesssim 0.04$, indicating that the disk is radiatively inefficient. Figure~\ref{moutplt} depicts that the mass outflow rate decreases with time rapidly whereas the mass accretion rate has little time variation. Considering that the ratio of mass accretion to outflow rates is given by $\dot{M}_{\rm out}/\dot{M}_{\rm a} = f_0 / (1 - f_0)$, we obtain the condition, by using equation (\ref{f0eqt}), that $\dot{M}_{\rm fb} < 5.5~\dot{M}_{\rm Edd}$ when $\dot{M}_{\rm out}$ is lower than $\dot{M}_{\rm a}$. Combining equations (\ref{mfb}) and (\ref{Medd}) find $t \sim 80~{\rm days}$ for $\dot{M}_{\rm fb} = 5.5~\dot{M}_{\rm Edd}$. The significantly low mass outflow rate at such late times makes the photosphere radius lower, resulting in lower thermal luminosity.

We have assumed that there is no dynamical interaction with ambient matter around the SMBH in our disk-wind model. More realistically, however, the interaction between the ambient mass and the outflow can occur and consequently create a shocked region from which non-thermal radiations, such as radio emission, are emitted. In fact, several radio-emitting TDEs, even without a relativistic jet, have been observed so far \citep{alexander_radio_2020}. To explore some mechanism for radio emission from radio-emitting, non-jetted TDE AT2019dsg \citep{cendes_radio_2021}, the theoretical work with a thin shell approximation has recently been done by \citet{2023ApJ...954....5H}. Constructing a unified model to reproduce observational thermal and non-thermal emissions from TDEs is also our future task.

The Doppler broadening of line emission arises from the motion of absorbing or emitting atoms, influenced by turbulence and thermal velocity which depends on gas temperature and atomic mass. Our CLOUDY modeling considers the density and temperature at the wind's inner radius, the helium-to-hydrogen ratio, and the filling factor as free parameters. Using these inputs, CLOUDY estimates gas temperature and ionization states, which affect emission line luminosity and broadening. While our CLOUDY modeling yields an FWHM for the H$\alpha$ line consistent with observations, the FWHM for the He II line is underestimated.
Those findings suggest that the discrepancy of the FWHM between the observation and CLOUDY modeling may originate from the assumption of steady wind gas motion. Realistically, the gas in the wind can accelerate due to the radially declining black hole gravity and the radiation force from photons. Therefore, we need to relax our steady-state assumption and apply time-dependent CLOUDY modeling to explore the FHWM discrepancy problem.


The center of spectral lines appears constant within the scatter for the first 90 days. The HeII line seems marginally blue-shifted with a velocity of $- 700 \pm 700~{\rm km~s^{-1}}$. However, this shift in line velocity lies within the FWHM. The H$\alpha$ line is consistent with the reference wavelength. This suggests the spectrum of iPTF16fnl is symmetric about the central wavelength of the spectral line. Our disk-wind model assumes a spherical distribution extending up to a large distance such that the wind outer radius $r_{\rm w,o} \gg r_{l}$ and the wind velocity is constant at all the radii. This simple assumption is appropriate for TDEs showing symmetric line profiles. In future work, we will consider a time-varying wind velocity and/or a non-spherical wind geometry to explain asymmetric spectral line profiles. 

\section{Summary}
\label{summary}
%
iPTF16fnl is a low redshift and optically low luminosity TDE with no significant X-ray observations, indicating the presence of an atmosphere that obscures the disk radiation and emits in UV/optical bands. The observed line signatures demonstrate that the Helium spectral lines dominate over the hydrogen lines at early times, and the luminosity ratio of helium to hydrogen lines decreases with time. Our paper aims to elucidate the physical properties of the atmosphere necessary for the observed spectrum continuum and line luminosity. We have estimated five unknown physical parameters by fitting the steady-state slim disk with a spherical wind model (i.e., disk-wind model) to the spectrum continuum at twelve observed time epochs. Moreover, we have studied why the helium line dominates over the hydrogen line by comparing the predicted lines by CLOUDY modeling with the observational spectral lines at three observed time epochs. Our key findings are summarized as follows:

\begin{enumerate}
\item
We find the black hole mass is $M_{\bullet} = 4.43^{+0.17}_{-0.14} \times 10^5 M_{\odot}$, the stellar mass is $M_{\star} = 0.46^{+0.12}_{-0.06} M_{\odot}$, the wind inner radius is $r_l = 1.88^{+0.293}_{-0.224} \times 10^{13}~{\rm cm}$, the wind velocity is $v_{\rm w}  = 5382.7^{+359.4}_{-347.1} ~{\rm km~s^{-1}}$, and the disk formation time is $\Delta t = 0.79^{+1.24}_{-0.572} ~{\rm days}$. Note that the time origin of our model is the time when debris with the most tightly bound orbit returns to the pericenter.

\item
In our disk-wind model, the radiative efficiency depends on the black hole mass and the mass accretion rate, indicating that the bolometric luminosity is not simply proportional to the mass accretion rate (see equation~\ref{etadisk}). We find the radiative efficiency to be $0.01\lesssim\eta\lesssim 0.04$ over the observational time, resulting in the disk being radiatively inefficient. The resultant emission dominates in the optical/UV, with a low X-ray luminosity of $\sim10^{40}\,{\rm erg\,s^{-1}}$.

\item
The mass outflow rate dominates the mass accretion rate at early times, but the mass outflow rate declines rapidly compared to the mass accretion rate. The significant decay of the mass outflow rate reduces the wind density so that the photosphere radius decreases with time. We confirm the photospheric radius agrees with the blackbody radius estimated by Blagorodonova (2017). 

\item
The photosphere temperature increases with time from $\sim 15000~{\rm K}$ to $\sim 50000~{\rm K}$. The peak bolometric luminosity at the photosphere ($\sim 6^{+3}_{-2} \times 10^{42}~{\rm erg~s^{-1}}$) is in good agreement with the observed OUV luminosity ($\sim 10^{43}~{\rm erg~s^{-1}}$), within the associated error limits. In contrast, the disk luminosity in soft-X-ray wavebands is low enough to support the dominance of optical/UV observations. 

\item
The CLOUDY modeling shows that the medium is generally inhomogeneous and clumpy. We estimate the filling factor to be 0.8, indicating that the gas occupies $80\%$ of the atmosphere as clumps while the remaining $20\%$ is a vacuum. 

\item
The CLOUDY modeling finds that the helium-to-hydrogen number ratio, $x$, lies between 0.1 and 0.15. The low value of $x$, which is comparable to $ x \sim 0.1$ for the sun, suggests the disrupted star originally is a main-sequence star. The CLOUDY modeling also demonstrates that the optical depth for the hydrogen emission line is two orders of magnitude higher than the helium emission line. Considering the helium abundance is $10\%$ to $15 \%$ of the hydrogen, we conclude that the dominance of the helium emission line over the hydrogen emission line is due to the optical depth effect.
\end{enumerate}

%
\section{acknowledgments}
%
We thank the referee for the constructive suggestions that have improved the paper. GS acknowledges WOS-A grant from the Department of Science and Technology (SR/WOS-A/PM-2/2021). Two authors have been supported by the Basic Science Research Program through the National Research Foundation of Korea (NRF) funded by the Ministry of Education (2016R1A5A1013277 to K.H. and 2020R1A2C1007219 to K.H. and M.T.)

%
\section{Data Availability}
%
Simulations in this paper made use of freely downloadable code CLOUDY (https://www.nublado.org, C17.02). The photometry data for the continuum is available at \url{https://cdsarc.cds.unistra.fr/viz-bin/cat/J/ApJ/844/46\#/browse}. The spectroscopy data is available at \url{https://www.wiserep.org/object/1710}.
The model-generated data are available on request.
\appendix

%
\section{Disk wind model details}
\label{dwmodel_theory}
%

In this section, we present the disk-wind model in detail. 
We consider a steady-state, radiation-pressure dominant, slim dis model with a super-Eddington accretion rate. The strong radiation pressure induces an outflow \citep{2009MNRAS.400.2070S,2015mnras.448.3514c,2019ApJ...885...93F}. In the slim disk model, the advection energy cooling is essential for the thermal stability of the disk with radiation pressure, reducing the amount of viscous heating flux radiated and the scale height of a slim disk is $H \sim R$ \citep{2002apa..book.....F,2009MNRAS.400.2070S}, where $R$ is the disk radius at the disk mid-plane. In a steady-state super-Eddington disk with an outflow, the mass conservation law gives equations (\ref{eq:mdotout}) and (\ref{eq:mdotacc}). These equations represent $f_0$ as a key parameter to decide the evolution of both rates. \citet{2011MNRAS.413.1623D} numerically simulated $f_0$ at the outer radius for a steady disk and showed that $f_0$ is approximated into empirical relation shown as equation (\ref{f0eqt}). Although \citet{2009MNRAS.400.2070S} assumed $f_0$ to be a constant, our model demonstrates that $f_0$ is not a constant because $\eta$ is also not constant, as seen in equations~(\ref{f0eqt}) and (\ref{etadisk}). The radiative flux from a super-Eddington disk with radiation pressure is given by \citep{2009MNRAS.400.2070S}  as

\begin{multline}
\sigma T_{\rm eff}^4 = \frac{3 G M_{\bullet} \dot{M}_{\rm a} f_R}{8 \pi R^3} \times \\  \left[\frac{1}{2}+ \sqrt{\frac{1}{4}+ \frac{3}{2} f_R 
\left(\frac{\dot{M}_{\rm a}}{L_{\rm Edd}/c^2}\right)^{2} 
\left(\frac{R}{R_{\rm S}}\right)^{-2}}\right]^{-1},
\label{tefdisc}
\end{multline}
where $\sigma$ is the Stefan-Boltzmann constant,
$R_{\rm S} = 2 G M_{\bullet}/c^2$ is the Schwarzschild radius, $f_R = 1 - \sqrt{R_{\rm in} / R}$, and $R_{\rm in}$ is the disk inner radius taken to be innermost stable circular orbit of non-spinning black hole, i.e. $R_{\rm in} = 3 R_{\rm S}$. A similar formula as the above equation for the radiative to advection cooling rate for a general relativistic slim disk model with radiation pressure was constructed by \citet{2023PhRvD.108d3021M}. 
The observed flux from the disk is given by \citep{2002apa..book.....F}
\begin{equation}
F =  \frac{\cos\theta_{\rm los}}{D_{\rm L}^2} \int_{\nu_{\rm l}}^{\nu_{\rm h}} \int_{R_{\rm in}}^{R_{\rm out}} B(T_{\rm eff},~\nu) 2 \pi R \, \diff R \, \diff \nu,
\label{eq:flux}
\end{equation}
where $R_{\rm out}$ is the disk outer radius, $B(T_{\rm eff},~\nu)$ is the blackbody Planck function, $\nu$ is the frequency, $\nu_{\rm l}$ and $\nu_{\rm h}$ are the lower and upper frequency respectively, $D_{\rm L}$ is the luminosity distance of the source to the observer and $\theta_{\rm los}$ is the is the angle between observer line-of-sight and disk normal vector. 
Adopting $\theta_{\rm los} = 0^{\circ}$ for equation~(\ref{eq:flux}), we obtain the observed luminosity as
\begin{eqnarray}
L 
&=&
4 \pi D_{\rm L}^2 F
\nonumber \\
 &=&
8 \pi^2 
 \int_{\nu_{\rm l}}^{\nu_{\rm h}} \int_{R_{\rm in}}^{R_{\rm out}} B(T_{\rm eff},~\nu) R \, \diff R \, \diff \nu.
 \nonumber 
\end{eqnarray}
The bolometric luminosity is obtained by applying $\nu_{\rm l} = 0$ and $\nu_{\rm h} = \infty$ to the above equation as
\begin{equation}
L_{\rm b} = 8 \pi \int_{R_{\rm in}}^{R_{\rm out}} \sigma T_{\rm eff}^4 R \, \diff R.
\label{lumb}
\end{equation}
The X-ray luminosity with the energy range 0.3-10 keV is estimated to be
\begin{equation}
L_{\rm X} = 8 \pi^2 \cos\theta_{\rm los} \int_{\nu_{\rm l,X}}^{\nu_{\rm h,X}} \int_{R_{\rm in}}^{R_{\rm out}} B(T_{\rm eff},~\nu) R \, \diff R \, \diff \nu,
\label{xlum}
\end{equation}
where the corresponding frequency range is from $\nu_{\rm l,X}=7.2 \times 10^{16}\,{\rm Hz}$ to $\nu_{\rm h,X}=2.4 \times 10^{18}$ Hz.

We assume that the wind is blowing spherically from the disk with the constant velocity $v_{\rm w}$ and mass outflow rate $\dot{M}_{\rm out}$. The radial density profile of the wind is then given by 
\begin{equation}
\rho(r) = \frac{\dot{M}_{\rm out}}{4 \pi r^2 v_{\rm w}},
\label{wden}
\end{equation}
where $\dot{M}_{\rm out} = f_0 \dot{M}_{\rm fb}$ (see equation~\ref{eq:mdotout}). We also assume that the wind is launched at the inner radius of the wind, $r_{\rm l}$ and the outer radius of the wind $r_{\rm w,o}$ is much larger than the inner radius, i.e., $r_{\rm w,o} \gg r_{l}$. The optical depth $\tau$ with an opacity dominated by electron scattering is given by 
\begin{eqnarray}
\tau_{\rm es}(r) 
&=& 
\int_{r}^{r_{\rm w,o}} \kappa_{\rm es} \rho(r) \, \diff r 
\approx
\frac{\kappa_{\rm es} \dot{M}_{\rm out} }{ 4 \pi v_{\rm w} r}.
\nonumber\\
&=& 4.45 \left(\frac{M_{\bullet}}{10^6 M_{\odot}}\right) \left(\frac{\eta}{0.01}\right)^{-1} \left(\frac{r}{10^{14}{\rm ~cm}}\right)^{-1} 
\nonumber \\
&\times&
\left(\frac{v_{\rm w}}{10^4 ~{\rm km~s^{-1}}}\right)^{-1} \left(\frac{\dot{M}_{\rm out}}{\dot{M}_{\rm Edd}}\right). 
 \label{eq:taues}
\end{eqnarray}
Considering the kinetic energy density of the wind, $(1/2)\rho(r) v_{\rm w}^2$, is in equilibrium with the internal energy density dominated by the radiation, $a T_l^4$, \citep{2009MNRAS.400.2070S}, we obtain the temperature $T_{l}$ at the inner radius of the wind as
\begin{equation}
T_l  = \left(\frac{\dot{M}_{\rm out}v_{\rm w}}{8 \pi r_{l}^2 a} \right)^{1/4},   
\label{eq:tl}
\end{equation}
where $a$ is the radiation constant.

If the photon diffusion time is longer than the dynamical time of the wind, the photons are trapped and coupled with the matter. 
By equating the photon diffusion timescale, $t_{\rm diff} = \tau_{\rm es}(r) r / c$, with the dynamical time of the wind, $t_{\rm dyn} = (r - r_l)/v_{\rm w}$, we obtain the photon trapping radius as
\begin{equation}
r_{\rm tr} = r_l + \frac{\kappa_{\rm es} \dot{M}_{\rm out}}{4 \pi c}.
\label{rtr}
\end{equation}
For $r < r_{\rm tr}$, the photons are advected without escaping outside so that the temperature evolves adiabatically.
Since the adiabatic temperature is given by $T \propto \rho(r)^{1/3} \propto r^{-2/3}$, the temperature at the trapping radius is given by 
\begin{equation}
T_{\rm tr} 
= 
T_l 
\left(
\frac{r_{\rm tr}}{r_l}
\right)^{-2/3}. 
\label{eq:tr}
\end{equation}
For $r > r_{\rm tr}$, the photons diffuse in the wind at a power:
\begin{equation}
L_{\rm w} = -\frac{4 \pi r^2 a c}{3 \kappa_{\rm es} \rho(r)} \frac{\partial T^4}{\partial r}.
\label{lumd}
\end{equation}

Considering the luminosity is nearly constant in the diffusive regime \citep{2020ApJ...897..156U}, the temperature evolves as $T \propto r^{-3/4}$. 
Substituting equations~(\ref{wden}) to~(\ref{eq:tr}) into equation~(\ref{lumd}) with $T\propto{r^{-3/4}}$ gives the luminosity at the trapping radius  as
\begin{equation}
L_{\rm ph} = \frac{2 \pi c}{\kappa_{\rm es}} v_{\rm w}^2 r_l \left(\frac{r_{\rm tr}}{r_l}\right)^{1/3}.
\label{lwind}
\end{equation}
In the diffusive region, apart from electron scattering, the absorption of photons by electrons can also be a source of opacity. Thus, total opacity is given by $\kappa_{\rm eff} \approx \sqrt{3 \kappa_{\rm es} \kappa_a}$, where the absorption opacity is taken to be the Kramers opacity, $\kappa_a = \kappa_0 \rho T^{-7/2} ~{\rm cm^{2}~g^{-1}}$. 
Such a color radius that the optical depth is unity, which is obtained by imposing $\tau_{\rm es}(r=r_{\rm cl})=1$ on equation~(\ref{eq:taues}),
is given by \citep{2020ApJ...897..156U}.
\begin{equation}
r_{\rm cl} = \left(\frac{16}{11}\right)^{16/11} \left(3 \kappa_{\rm es}\kappa_0\right)^{8/11} \left(\frac{\dot{M}_{\rm out}}{4 \pi v_{\rm w}}\right)^{24/11} r_{\rm tr}^{-21/11} T_{\rm tr}^{-28/11}.
\label{eq:rcl}
\end{equation}
The photospheric radius is given by
\begin{equation}
r_{\rm ph} = {\rm Max [} r_{\rm tr},~r_{\rm cl} {\rm]} 
\approx 
r_{\rm tr}
\label{eq:rph}
\end{equation}
for appropriate parameters. So, we confirm that equation~(\ref{lwind}) corresponds to the luminosity at the photosphere.
The photospheric temperature is then given by  
\begin{equation}
T_{\rm ph} = \left(\frac{L_{\rm ph}}{4 \pi r_{\rm ph}^2 \sigma}\right)^{1/4},
\label{eq:tph}
\end{equation}
where where we use equation~(\ref{lwind}) for the derivation.
Assuming a blackbody emission from the photosphere, 
we obtain the effective luminosity at a given frequency as \citep{2011MNRAS.415..168S}
\begin{equation}
L^{\rm w}(\nu)
=
\nu L_{\nu,\rm ph} = \nu ~4 \pi r_{\rm ph}^2  \frac{\pi 2 h \nu^3}{c^2} \left[ \exp\Big( \frac{h \nu}{k_B T_{\rm ph}} \Big)-1 \right]^{-1}
\label{lspec}
\end{equation}
with the spectral luminosity at the photosphere: 
\begin{equation}
L_{\nu,\rm ph} \equiv 4 \pi r_{\rm ph}^2  \frac{\pi 2 h \nu^3}{c^2} \left[ \exp\Big( \frac{h \nu}{k_B T_{\rm ph}} \Big)-1 \right]^{-1}.
\nonumber
\end{equation}
The X-ray luminosity at the photosphere in the range of $0.3-10$\,{\rm keV} is calculated by
\begin{eqnarray}
L_{X,\rm ph}
&=& 
\int_{\nu_{\rm l,X}}^{\nu_{\rm h,X}}
L_{\nu,\rm ph} 
\,d\nu
\nonumber \\
&=& 
\frac{8\pi^2 h r_{\rm ph}^2}{c^2}
\int_{\nu_{\rm l,X}}^{\nu_{\rm h,X}}
\nu^3
\left[ \exp\Big( \frac{h \nu}{k_B T_{\rm ph}} \Big)-1 \right]^{-1}
\,
d\nu,
\nonumber \\
\label{eq:lx_rph}
\end{eqnarray}
where the values of $\nu_{\rm l,X}$ and $\nu_{\rm h,X}$ are seen below equation (\ref{xlum}).

\bibliography{main}
\end{document}